\documentclass[11pt]{article}

\newsavebox{\PSLASH}
\sbox{\PSLASH}{$p$\hspace{-1.8mm}/}

\begin{document}
\title{Some Aspects of $c$$=-2$ Theory }
\author{M. A. Rajabpour\footnote{e-mail: rajabpour@physics.sharif.ir}
, S. Rouhani \footnote{e-mail: rouhani@ipm.ir}  and A. A. Saberi \footnote{e-mail: a$_{_{-}}$saberi@physics.sharif.ir} \\ \\
Department of Physics, Sharif University of Technology,\\ Tehran,
P.O.Box: 11365-9161, Iran} \maketitle
\begin{abstract}
We investigate some aspects of the c=-2 logarithmic conformal
field theory. These include the various representations related
to this theory, the structures which come out of the Zhu algebra
and the W algebra related to this theory. We try to find the
fermionic representations of all of the fields in the extended
Kac table especially for the untwisted sector case. In addition,
we calculate the various OPEs of the fields, especially the
energy-momentum tensor. Moreover, we investigate the important
role of the zero modes in this model. We close the paper by
considering the perturbations of this theory and their
relationship to integrable models and generalization of
Zamolodchikov's $c-$theorem.
  \vspace{5mm}%
\end{abstract}
\section{Introduction}\

Conformal field theory (CFT) as a powerful tool for investigating
critical systems initially appeared in the seminal paper by
Belavin, Polyakov and Zamolodchikov \cite{bpz}. Following this
paper many other papers appeared expanding various aspects of
CFT, one may look up the books \cite{Ds,H} for good introduction
to the field. Conformal field theory investigates the quantum
theory of models with conformal invariance. However it is most
powerful in two dimensions where the conformal group is infinite
dimensional. It also turns out that in two dimensions, systems
which have scale invariance are also conformally invariant,
though there is a counter example to this rule \cite{rc}. Soon
interest was shown in theories with larger symmetries such as
Wess, Zummino, Witten, Novikov models \cite{wzwn}. The other way
to extend the conformal symmetry is the introduction of the $W$
algebra, firstly investigated by Zamolodchikov \cite{zamol}. This
closed algebra is a generalization of Virasoro algebra by
introducing some primary operators, for a good review see
\cite{schutens}.

 The advent of CFT was in tune with another development namely
 the development of integrable models in field theory \cite{zz} and
 exactly solvable models in statistical mechanics \cite{baxter}. In
 integrable models one also has an infinite number of conserved
 currents;
 but unlike CFT, an infinite dimensional symmetry group is not
 involved. Therefore one may view CFT as a special integrable
 model, in fact CFT describes
  fixed points in the renormalization flow of integrable
 models\cite{zamol2}.
 Therefore it is natural to investigate integrable models as perturbations of CFTs, this
 was first addressed by Zamolodchikov \cite{zamolo}. In this paper
 Zamolodchikov solved the two dimensional Ising model in the presence of a magnetic
 field at the critical point. After Zamolodchikov, many papers in
 this direction appeared, for example see \cite{mussardo,delfino}.

 The other important version of CFT is Logarithmic CFT (LCFT) which contains
 unusual representations of Virasoro algebra. LCFT first appeared in
 the works of  Kniznick \cite{kniz} and then  Saluer and Rozanskey \cite{s,sr}. But
 LCFT as a new representation of Virasoro algebra with some
 special operator product expansion(OPE) was studied first by
 Gurarie \cite{gurarie}. He showed that LCFT contains an irreducible but
 indecomposable representation of Virasoro algebra and as an
 example he studied the $c=-2$ model. The $c=-2$ action was first
 introduced in \cite{dens} for studying dense polymers. After the work
 of Gurarie many works were done to investigate various aspects of $c=-2$ model.
 The $W(2,3)$ algebra structure was first discovered by Kausch \cite{k1}.
 Some other $W$ algebra structures of $c=-2$ model were established
 by other authors like \cite{eh}. In this paper all
 of the highest weight representations of the $c=-2$ model were
 introduced for the first time. This work was completed by the Gaberdiel and Kausch  in
 the series of papers \cite{k2,kg1,kg2,kg3,k3}. All of important results are gathered in
 Gaberdiel's review paper \cite{g}.The $c=-2$ model is also interesting for
 its many applications, some of which are dense
 polymers \cite{s,dens}, sandpile models \cite{ruell,sand,jpr}, Quantum hall effect \cite{f1,kogan}
 and dressing quantum gravity \cite{kl}, And recently new
 logarithmic integrable lattice models were introduced which is related
 to the critical dense polymers \cite{pr,prz}.
 Moreover the boundary $c=-2$ was studied widely by many authors
 \cite{boundaryc=-2} which many of the results are gathered in
 \cite{boundary gaberdiel}.
 Because of its many applications together with being the simplest example of LCFTs,
 it is widely believed that $c=-2$ is a good lab to study LCFTs.\\

 In this work we want to investigate many different aspects of
 $c=-2$ model. Our works are based on the free ghost action. We
 start by introducing the action and then identifying some of
 the concepts derived for LCFTs before in the context of this action and the fields inside the
 model.\\
The paper is organized as follows: In the second section we review
most important properties of LCFTs, it is a brief review with some
new results, for more details you can see the classic reviews
\cite{Flohr,moghimi}. In the third section, by introducing the
$c=-2$ action, we investigate some primary fields as well as the
logarithmic partners. In this section we calculate some useful
correlation functions and OPEs which are in agreement with the
second section's results. In addition we establish the algebraic
method of Gaberdiel and Kausch  for classification of $c=-2$
representations. This is based on the series of papers
\cite{k2,kg1,kg2,kg3,k3,g}. We relate these representations to the
fields which we investigate at the beginning of the third section.
In the forth section we calculate correlation functions of
logarithmic energy-momentum tensor with some other fields like
stress-energy tensor and those fields which appear in the OPE of
these two fields. In addition we establish the logarithmic
Sugawara construction of Kogan and Nichols \cite{kn}. In the
fifth section, we will study the extended Kac table of model and
try to relate every elements of the grid to the known fields
defined in the previous sections. In addition we present some
facts about the $W$ algebra structure of $c=-2$ model. This
structure of $c=-2$ is not known well so far.

In the sixth section, we study some different ways to perturb the
model and try to find some of perturbations which are related to
integrable models. We concentrate on finding the conserved
quantities, some of the results have appeared in our previous work
\cite{rr}.

Finally, we investigate some aspects of Zamolodchikov's $c$
theorem in the $c=-2$ model and see how the zero mode can affect
the structure of this theorem. Moreover, we study the generalized
$c$ theorem, which is definable in the integrable models, and find
the generalized renormalization group (RG) quantities which
decrease under the RG flow.

\section{Logarithmic CFT Theory}\
\setcounter{equation}{0}

An indicative feature of LCFT models is that there are  a number
of primary fields with the same conformal weights. Clearly the
action of the stress energy tensor on such a set is not
diagonalisable.  As an example let's take a pair of operators as
$\Phi$ and $\Psi$ which have the following OPE  with the
energy-momentum tensor:

\begin{eqnarray}\label{T and t1}
 T(z)\Phi(w)=\frac{h\Phi(w)}{(z-w)^2}+\frac{\partial\Phi(w)}{z-w}+\cdot\cdot\cdot
\end{eqnarray}
\begin{eqnarray}\label{T and t2}
T(z)\Psi(w)=\frac{h\Psi+\Phi(w)}{(z-w)^2}+\frac{\partial\Psi(w)}{z-w}+\cdot\cdot\cdot.
\end{eqnarray}
One can see from above that the action of $L_{0}$ on the pair
$\Phi$, $\Psi$ is not diagonal rather it has a Jordan form. This
fact implies many interesting results, such as the appearance of
logarithms in correlation functions.

To investigate some of these results we begin by looking at the
infinitesimal transformations consistent with the above :

\begin{eqnarray}\label{infinitesimal1}
 \delta_{\epsilon}\Phi(z)&=&(h\partial_{z}\epsilon+\epsilon
 \partial_{z})\Phi(z)\\
\delta_{\epsilon}\Psi(z)&=&(h\partial_{z}\epsilon+\epsilon
 \partial_{z})\Psi(z)+\partial_{z}\epsilon \Phi(z).
\end{eqnarray}
One can rewrite the above equations in a compact form by using
nilpotent variables \cite{moghimi}

\begin{eqnarray}\label{infinitesimal2}
 \delta_{\epsilon}\Phi(z,\lambda)=\left((h+\lambda)\partial_{z}\epsilon+\epsilon
 \partial_{z}\right)\Phi(z,\lambda).
\end{eqnarray}
where $\lambda^{2}=0$ and $\Phi(z,\lambda)=\Phi(z)+\lambda \Psi(z)$,
for the bigger Jordan cells it is sufficient to assume $\lambda
^{n}=0$, and expand $\Phi(z,\lambda)$ accordingly . In this article
we work with just rank two cells, hence we shall take
$\lambda^{2}=0$. This implies that conformal weights have a
nilpotent component and we may derive the finite transformations
under a conformal mapping of the complex plane $\textit{w}$ as:
\begin{eqnarray}\label{finite}
 \Phi(z,\lambda)=\left(\frac{\partial\textit{w}}{\partial\textit{z}}\right)^{h+\lambda}\Phi(\textit{w},\lambda)
\end{eqnarray}

Now we can derive the two point functions of the pair of
logarithmic fields by using the (\ref{finite}), invariance under
the translation, rotation, scale and special conformal
transformations
\begin{eqnarray}\label{two point}
 \langle\Phi(z,\lambda_{1})\Phi(w,\lambda_{2})\rangle=\frac{a(\lambda_{1},\lambda_{2})}{(z-w)^{2h+\lambda_{1}+\lambda_{2}}},
\end{eqnarray}
where
$a(\lambda_{1},\lambda_{2})=a_{1}(\lambda_{1}+\lambda_{2})+a_{12}\lambda_{1}\lambda_{2}$.
The interesting result is that the two point function of the field
$\Phi(z)$ is zero and there is logarithm in the two point function
of the field $\Psi(z)$ which is named the logarithmic partner of
the field $\Phi(z)$. Similar results can be derived for the
higher order correlators. For example the three point functions
of the logarithmic fields can be derived by using equation
(\ref{finite}) like two point functions; they have the following
form:
\begin{eqnarray}\label{three point}
 \langle\Phi(z_{1},\lambda_{1})\Phi(z_{2},\lambda_{2})\Phi(z_{3},\lambda_{3})\rangle=
 f(\lambda_{1},\lambda_{2},\lambda_{3})z_{12}^{-a_{12}}z_{23}^{-a_{23}}z_{31}^{-a_{31}},
\end{eqnarray}
where $z_{ij}=(z_{i}-z_{j})$,
$a_{ij}=h_{i}+h_{j}-h_{k}+(\lambda_{i}+\lambda_{j}-\lambda_{k})$
and the $f(\lambda_{1},\lambda_{2},\lambda_{3})$ has the following
form:

\begin{eqnarray}\label{f}
f(\lambda_{1},\lambda_{2},\lambda_{3})=\sum_{i=1}^{3}C_{i}\lambda_{i}+\sum_{1\leq
i<j\leq3}C_{ij}\lambda_{i}\lambda_{j}+C_{123}\lambda_{1}\lambda_{2}\lambda_{3}.
\end{eqnarray}

Notice that the three point function of the field $\Phi(z)$ is
also zero like the two point function, one can prove this by using
the Ward identity. In general the $n$ point function of this field
is zero. A well known example is the identity operators in the
$c=-2$ model which we investigate in the next section. Note that
equation (\ref{three point}) actually contains four different
correlations.

One can extend the OPE (\ref{T and t2}) by inserting another
singular term on the rhs of the OPE of the energy-momentum tensor
with $\Psi$, for example suppose we have the following more
singular OPE:

\begin{eqnarray}\label{T and pesi}
 T(z)\Psi(w)=\frac{\xi(w)}{(z-w)^{3}}+\frac{h\Psi(w)+\Phi(w)}{(z-w)^2}+\frac{\partial\Psi(w)}{z-w}+\cdot\cdot\cdot
\end{eqnarray}
where $\xi$ is an ordinary (non logarithmic) primary field with
the conformal weight $h-1$. So one can derive the finite
transformation of $\Psi$ as:

\begin{eqnarray}\label{delta pesi}
 \Psi(z)=\left(\frac{\partial\textit{w}}{\partial\textit{z}}\right)^{h}\left(\Psi(w)+\log\left(\frac{\partial\textit{w}}{\partial\textit{z}}\right)
 \Phi(w)+\frac{\frac{\partial^{2}\textit{w}}{\partial^{2}\textit{z}}}{2\left(\frac{\partial\textit{w}}{\partial\textit{z}}\right)^{2}}\xi(w)\right).
\end{eqnarray}

We can check that the above equation is consistent under sequence
of successive analytic transformations $z\rightarrow w\rightarrow
w'$. One can now follow the familiar procedures to derive the
modified two point functions resulting out of the presence of the
last term $\xi(w)$. This term is not important in the translation,
rotation and scale transformation but the special conformal
transformation has to be taken into consideration. We can see that
the two point functions of $\Phi$ and $\Psi$ are similar to
(\ref{two point}), the difference arises in the two point function
of the $\xi$ and the other fields. We observe that the two point
function of $\xi$ with $\Phi$ is zero but the two point functions
of $\xi$ with itself and $\Psi$ have the following forms:

\begin{eqnarray}\label{kesi pesi}
 \langle\Psi(z)\xi(w)\rangle&=&\frac{b}{(z-w)^{2h-1}}\\
 \langle\xi(z)\xi(w)\rangle&=&\frac{b}{(z-w)^{2h-2}}.
\end{eqnarray}
Observe that $\xi$ and $\Psi$ have different conformal weights but
the two point functions are non vanishing. One can repeat the
above process for three point functions and observe that again the
three point functions of $\Phi$ and $\Psi$ remain unchanged
whereas the three point functions involving $\xi$ have the
following forms:

\begin{eqnarray}\label{3kesi pesi}
\langle\Phi(z_{1},\lambda_{1})\Phi(z_{2},\lambda_{2})\xi(z_{3})\rangle&=&
a\lambda_{1}\lambda_{2}z_{12}^{h+1}z_{23}^{h-1}z_{31}^{h-1}\\
\langle\Phi(z_{1},\lambda_{1})\xi(z_{2})\xi(z_{3})\rangle&=&c\lambda_{1}
z_{12}^{h}z_{23}^{h-2}z_{31}^{h}\\
\langle\xi(z_{1})\xi(z_{2})\xi(z_{3})\rangle&=&dz_{12}^{h-1}z_{23}^{h-1}z_{31}^{h-1}.
\end{eqnarray}

In the next section we give examples of these operators in the
$c=-2$ model and we establish some representations by using the
explicit fermionic action. It may happen that even more singular
fields appear in the OPE of the field with the energy-momentum
tensor. For such a case the correlators of these fields may also
be calculated similar to the above.

\section{The Logarithmic $c = -2$ Theory}\label{Logarithm}\
\setcounter{equation}{0}

The conventional $c =-2$ theory is constructed using a pair of free
grassmanian scalar fields: $\theta^{\alpha}=(\theta,\bar{\theta})$
with the
action, \\

\begin{equation}\label{s}
 S=\frac{1}{2\pi}\int\varepsilon_{\alpha\beta}\partial\theta^{\alpha}\bar{{\partial}}\theta^{\beta}
    =\frac{1}{\pi}\int\partial\theta\bar{\partial}\bar{\theta},
\end{equation}
where $\varepsilon$ is the canonical symplectic form,
$\varepsilon_{12}=+1$. Note that $\theta^{1}=\theta$ and $\theta^{2}=\bar{\theta}$.\\
In order to calculate the correlators we need to be careful about
the zero modes. Expanding in terms of modes we have:

\begin{equation}\label{te}
\theta^{\alpha}(z)=\sum_{n\neq0}\theta_{n}^{\alpha}z^{-n}+\theta_{0}^{\alpha}\log(z)+\xi^{\alpha}.
\end{equation}
Here $n$ is an integer number for untwisted sector which is
related to the periodic boundary condition and for the twisted
sector we must choose $n$ as half integer. The zero modes $\xi$
(and $\bar{\xi}$) appear which do not enter the action (\ref{s}).
So, to avoid the vanishing of any correlation function involving
$\theta^{\alpha}=(\theta,\bar{\theta})$, we have to insert these
zero modes in the expectations. Therefore, we can compute the
different nonzero correlation functions of $\theta$ and
$\bar{\theta}$

\begin{equation}\label{tete}
\langle\theta^{\alpha}(z)\theta^{\beta}(w)\bar{\xi}\xi\rangle= \varepsilon^{\alpha\beta}\log|z-w|  \\
\end{equation}
where $\varepsilon^{\alpha\beta}=-\varepsilon_{\alpha\beta}$. The
correlation functions related to the derivatives of $\theta$ and
$\bar{\theta}$ are simply obtained from the above relation for
instance

\begin{equation}\label{dtedte}
\langle\partial\theta^{\alpha}(z)\partial\theta^{\beta}(w)\bar{\xi}\xi\rangle= \varepsilon^{\alpha\beta}\frac{1}{2(z-w)^2}. \\
\end{equation}
Moreover, in this theory we have :

\begin{eqnarray}\label{tete1}
 \langle\bar{\theta}(w)\theta(z)\rangle=\langle\bar{\xi}\xi\rangle=1, \hspace{1cm}
 \langle1\rangle=0.
\end{eqnarray}

Let us now proceed to calculate the OPE of the energy-momentum
tensor $T=2:\partial\theta\partial\bar{\theta}:$ with itself. This
results in the characteristic OPE of a CFT with central charge
$c=-2$ :

\begin{eqnarray}\label{TT}
 T(z)T(w)=-\frac{1}{(z-w)^4}+\frac{2T(w)}{(z-w)^2}+\frac{\partial
 T(w)}{z-w}+\cdot\cdot\cdot.
\end{eqnarray}
Thus if we expand the energy-momentum tensor by using the modes we
have the following Virasoro algebra
\begin{eqnarray}\label{virasoro}
 [L_{n},L_{m}]=(n-m)L_{n+m}-\frac{1}{6}n(n^{2}-1)\delta_{n+m}.
\end{eqnarray}
 The same result can be found for the conjugate algebra
 $\bar{L}_{n}$ related to the $\bar{T}$.\\

 We continue this section by establishing some famous representations of
 $c=-2$ model such as $\mathcal{R}$$_{0}$, $\mathcal{R}$$_{1}$ and
 $\mathcal{R}$. The two representations $\mathcal{R}$$_{0}$, $\mathcal{R}$$_{1}$ are
 the highest weight representations but $\mathcal{R}$ is a local representation
 whose amplitudes are local. In the end of this section  we shall
 investigate briefly
 all of the highest weight representations of $c=-2$ model by using Zhu's method \cite{g}.

\subsection{$\mathcal{R}$$_{0}$ Representation}\

The $\mathcal{R}$$_{0}$ representation consists of the identity
operator and the field $:\theta\bar{\theta}:$ which is constructed
out of the elementary fields $\theta$ and $\bar{\theta}$. The
fields $\theta$ and $\bar{\theta}$ are primary fields with
conformal dimensions (0,0), while the bosonic composit operator
$:\theta\bar{\theta}:$ has the following OPE with $T$ :
\begin{eqnarray}\label{Ttete}
 T(z):\theta\bar{\theta}:(w)=-\frac{I}{2(z-w)^2}+\frac{\partial:\theta\bar{\theta}:(w)}{z-w}+\cdot\cdot\cdot.
\end{eqnarray}
If one rescales the operator $:\theta\bar{\theta}:$ as
$\tilde{I}=-2:\theta\bar{\theta}:$ in the above OPE, the
conventional logarithmic form is obtained. The field $\tilde{I}$
is the logarithmic partner of the identity and it is not an
ordinary primary field. A simple calculation shows that the fields
$\theta$ and $\bar{\theta}$ have zero conformal dimensions too.
The other zero dimension doublet is $\bar{\partial}\theta$ and
$\partial\bar{\theta}$.\\ In this way, representations of the
$c=-2$ theory can be constructed where there are many primary
fields. However in this subsection we investigate just the
simplest primary fields and in the forthcoming sections we
establish more complex primary fields and relate them to the
extended Kac table.

So one can obtain the OPEs of the above set of operators as
\cite{k2}:

\begin{eqnarray}\label{tetateta}
 \theta^{\alpha}(z)\theta^{\beta}(w)&=&\varepsilon^{\alpha\beta}(\frac{\tilde{I}(w)}{2}+\log|z-w|I)+...\\
\theta^{\alpha}(z)\tilde{I}(w)&=&-\log|z-w|^{2}\theta^{\alpha}(w)\\
\tilde{I}(z)\tilde{I}(w)&=&-\log|z-w|^{2}(\tilde{I}(w)+\log|z-w|^{2}I).
\end{eqnarray}

It is easy to check that the two point functions of the fields $I$
and $\tilde{I}$ satisfy the equation (\ref{two
 point}). We finish this subsection with the statement that $\mathcal{R}_{0}$
 representation is generated from a logarithmic pair of operators $I$ and
 $\tilde{I}$ with a Jordan cell structure of rank 2.

\subsection{$\mathcal{R}$$_{1}$ Representation}\

The $\mathcal{R}$$_{1}$ representation consists of the fields
$\varphi^{\alpha}=\partial\theta^{\alpha}$ and
$\psi^{\alpha}=:\partial\theta^{\alpha} \tilde{I}:$. The operators
$\varphi^{\alpha}$ have conformal weight $(1,0)$ and they are a
doublet with trivial correlation functions however the fields
$\psi^{\alpha}$ have unsual properties. The OPE of these fields
with $T$ has the following form:
\begin{eqnarray}\label{Tteta2}
 T(z)\psi^{\alpha}(w)=-\frac{\theta^{\alpha}(w)}{2(z-w)^3}+\frac{\varphi^{\alpha}(w)
 +\psi^{\alpha}(w)}{(z-w)^{2}}+
 \frac{\partial\psi^{\alpha}(w)}{z-w}+\cdot\cdot\cdot.
\end{eqnarray}
This is an example of (\ref{T and pesi}) with
$\xi(w)=-\frac{1}{2}\theta^{\alpha}(w)$. Many consequences follow,
as an example consider the following relations for the fields
$\psi^{\alpha}$ and $\varphi^{\alpha}$,
\begin{eqnarray}\label{lopration}
 L_{0}\varphi^{\alpha}=\varphi^{\alpha},
 \hspace{2cm}L_{0}\psi^{\alpha}=\psi^{\alpha}+\varphi^{\alpha},\hspace{2cm}L_{1}\psi^{\alpha}=-\frac{1}{2}\theta^{\alpha}.
\end{eqnarray}

One can see that $\varphi^{\alpha}$ and $\psi^{\alpha}$ form a
Jordan cell with respect to $L_0$, but $L_{1}$ does not vanish on
$\psi^{\alpha}$ , in fact this triplet of fields are related by the
action of the Virasoro generators.

\subsection{$\mathcal{R}$ Representation}\

The representation $\mathcal{R}$ is constructed out of $I$,
$\tilde{I}$ and the following higher level operators
\begin{eqnarray}\label{operatorr}
 \rho^{\alpha\bar{\alpha}}=\partial\theta^{\alpha}\theta^{\bar{\alpha}},\hspace{2cm}
 \bar{\rho}^{\alpha\bar{\alpha}}=-\bar{\partial}\theta^{\bar{\alpha}}\theta^{\alpha}
 \nonumber\\
 \varphi^{\alpha\bar{\alpha}}=\partial\theta^{\alpha}\bar{\partial}\theta^{\bar{\alpha}},\hspace{2cm}
\psi^{\alpha\bar{\alpha}}=:\varphi^{\alpha\bar{\alpha}}\tilde{I}:
\end{eqnarray}
The $\varphi^{\alpha\bar{\alpha}}$ are primary fields of weight
(1,1). The OPEs of $\rho^{\alpha\bar{\alpha}}$,
$\bar{\rho}^{\alpha\bar{\alpha}}$ and $\psi^{\alpha\bar{\alpha}}$
with energy-momentum tensor $T$  are given by
\begin{eqnarray}\label{OPEtro}
T(z)\rho^{\alpha\bar{\alpha}}(w)&=&\frac{\varepsilon^{\alpha\bar{\alpha}}I}{2(z-w)^{3}}+
\frac{\rho^{\alpha\bar{\alpha}}(w)}{(z-w)^{2}}+\frac{\partial\rho^{\alpha\bar{\alpha}}(w)}{z-w}+\cdot\cdot\cdot\\
T(z)\bar{\rho}^{\alpha\bar{\alpha}}(w)&=&
\frac{\varphi^{\alpha\bar{\alpha}}(w)}{z-w}+\cdot\cdot\cdot\\
T(z)\psi^{\alpha\bar{\alpha}}(w)&=&\frac{-\bar{\rho}^{\alpha\bar{\alpha}}(w)}{(z-w)^{3}}+
\frac{\varphi^{\alpha\bar{\alpha}(w)}+\psi^{\alpha\bar{\alpha}}(w)}{(z-w)^{2}}+
\frac{\partial\psi^{\alpha\bar{\alpha}}(w)}{z-w}+\cdot\cdot\cdot.\label{OPEtro2}
\end{eqnarray}
The OPE (\ref{OPEtro2}) is similar to the mentioned form (\ref{T
and pesi}), so one expects that the two point functions of
$\bar{\rho}^{\alpha\bar{\alpha}}$ with itself and
$\psi^{\alpha\bar{\alpha}}$ have the following forms as (\ref{two
point}) and (\ref{kesi pesi}):
\begin{eqnarray}\label{OPEtro}
\langle\bar{\rho}^{\alpha\bar{\alpha}}(z)\bar{\rho}^{\beta\bar{\beta}}(w)\rangle&=&
\frac{-\varepsilon^{\alpha\beta}\varepsilon^{\bar{\alpha}\bar{\beta}}}{2(\bar{z}-\bar{w})^{2}},\hspace{1.9cm}
\langle\bar{\rho}^{\alpha\bar{\alpha}}(z)\psi^{\beta\bar{\beta}}(w)\rangle=
\frac{\varepsilon^{\alpha\beta}\varepsilon^{\bar{\alpha}\bar{\beta}}}{2(\bar{z}-\bar{w})^{2}(z-w)},\nonumber\\
\langle\psi^{\alpha\bar{\alpha}}(z)\psi^{\beta\bar{\beta}}(w)\rangle&=&
\frac{\varepsilon^{\alpha\beta}\varepsilon^{\bar{\alpha}\bar{\beta}}}{2|z-w|^{4}}(1+\log|z-w|^{2}).
\end{eqnarray}
In addition, the other correlations can be obtained as bellow :
\begin{eqnarray}\label{two point3}
\langle\tilde{I}(z)\rho^{\alpha\bar{\alpha}}(w)\rangle&=&\frac{\varepsilon^{\alpha\bar{\alpha}}}{z-w},\hspace{1.6cm}
\langle\tilde{I}(z)\tilde{\rho}^{\alpha\bar{\alpha}}(w)\rangle=\frac{\varepsilon^{\alpha\bar{\alpha}}}{\bar{z}-\bar{w}},\nonumber\\
\langle\tilde{I}(z)\psi^{\alpha\bar{\alpha}}(w)\rangle&=&\frac{-\varepsilon^{\alpha\bar{\alpha}}}{|z-w|^{2}},\hspace{.9cm}
\langle\rho^{\alpha\bar{\alpha}}(z)\rho^{\beta\bar{\beta}}(w)\rangle=
\frac{-\varepsilon^{\alpha\beta}\varepsilon^{\bar{\alpha}\bar{\beta}}}{2(z-w)^{2}},\nonumber\\
\langle\rho^{\alpha\bar{\alpha}}(z)\bar{\rho}^{\beta\bar{\beta}}(w)\rangle&=&0,\hspace{2cm}
\langle\rho^{\alpha\bar{\alpha}}(z)\psi^{\beta\bar{\beta}}(w)\rangle=
\frac{\varepsilon^{\alpha\beta}\varepsilon^{\bar{\alpha}\bar{\beta}}}{2(z-w)^{2}(\bar{z}-\bar{w})},\nonumber\\
\langle\rho^{\alpha\bar{\alpha}}(z)\varphi^{\beta\bar{\beta}}(w)\rangle&=&0,\hspace{2.2cm}
\langle\psi^{\alpha\bar{\alpha}}(z)\varphi^{\beta\bar{\beta}}(w)\rangle=
\frac{-\varepsilon^{\alpha\beta}\varepsilon^{\bar{\alpha}\bar{\beta}}}{2|z-w|^{4}},\nonumber\\
\langle\varphi^{\alpha\bar{\alpha}}(z)\varphi^{\beta\bar{\beta}}(w)\rangle&=&0.
\end{eqnarray}
 The fields $\varphi^{\alpha\bar{\alpha}}$ for
 $\alpha\neq\bar{\alpha}$ are the same as those which appear in the
 action (\ref{action4}). In section ($6$) we show that if one
 perturbs the action with the fields $\varphi^{\alpha\alpha}$ then
 the theory remains conformal.

\subsection{Fields with unusual Logarithmic OPEs}\

One of the important fields in the $c=-2$ theory that is well
known in sandpile model \cite{ruell,rr} is
$\phi=:\partial\theta\bar{\partial}\bar{\theta}+\bar{\partial}\theta\partial\bar{\theta}:$
which is composed of the $\mathcal{R}$ representation's fields.
These fields are related to some height correlation functions in
sandpile model playing a role in spanning trees and spanning
forest \cite{forest}. Using the OPE of this field with
energy-momentum tensor one can simply read off that $\phi$ is a
primary field with conformal dimensions $(1,1)$

\begin{eqnarray}\label{Tfi}
 T(z)\phi(w)=\frac{\phi(w)}{(z-w)^2}+\frac{\partial\phi(w)}{z-w}+\cdot\cdot\cdot.
\end{eqnarray}
Also $\phi$ is a descendent of $:\theta\bar{\theta}:$ and the
Virasoro algebra generators $L_n$'s operate on this field as

\begin{eqnarray}\label{Lnphi}
L_{-1}\phi=\partial \phi, \hspace{2cm} L_0\phi=\phi\hspace{1cm} and
\hspace{1cm} L_n\phi=0\hspace{0.5cm}for\hspace{0.5cm} n\geq1.
\end{eqnarray}
The OPE of $\phi$ with itself has the following form :
\begin{eqnarray}\label{fifi}
 \phi(z)\phi(w)=-\frac{I}{2|z-w|^4}+\frac{\bar{T}(w)}{2(z-w)^2}
 +\frac{T(w)}{2(\bar{z}-\bar{w})^2}+\cdot\cdot\cdot.
\end{eqnarray}

So we can write the closed algebra for The modes of $\phi$. If we
show the Modes as $\phi_{m,n}$ then we can write the algebra as
the following form by using the Virasoro modes

\begin{eqnarray}\label{mode expansion}
[\phi_{m,n},\phi_{p,q}]&=&\frac{nm}{4}\delta_{n+q,0}\delta_{m+p,0}
 +\frac{m}{2}\delta_{m+p,0}L_{n+q-2}+\frac{n}{2}\delta_{n+q,0}\bar{L}_{m+p-2}.
\end{eqnarray}
In addition the OPE of $\phi$ with the $:\theta\bar{\theta}:$ is

\begin{eqnarray}\label{fitete}
 \phi(z):\theta\bar{\theta}:(w)=-\frac{I}{2|z-w|^2}+\frac{\bar{\partial}:\theta\bar{\theta}:(w)}{2(z-w)}+
 \frac{\partial:\theta\bar{\theta}:(w)}{2(\bar{z}-\bar{w})}+\cdot\cdot\cdot.
\end{eqnarray}
By using the above OPE one can write a closed algebra for the
modes of $\phi$ and $\tilde{I}$ too. Insertion of
$:\theta\bar{\theta}:$ aside $\phi$ yields a logarithmic field
$\psi=:\partial\theta\bar{\partial}\bar{\theta}\theta\bar{\theta}
+\bar{\partial}\theta\partial\bar{\theta}\theta\bar{\theta}:$.

 It is believed that the field $\psi$ is a logarithmic
partner of $\phi$, and our explicit calculation of OPE of $\psi$
with the energy-momentum tensor shows that too

\begin{eqnarray}\label{Tsi}
 T(z)\psi(w)=\frac{\bar{\partial}:\theta\bar{\theta}:}{2(z-w)^3}+\frac{\psi(w)-\frac{1}{2}\phi(w)}{(z-w)^2}
 +\frac{\partial\psi(w)}{(z-w)}+\cdot\cdot\cdot.
\end{eqnarray}
In this expression the third order singular term appears whereas
the standard form of OPEs in LCFT has no similar term. The field
$\frac{1}{2}\bar{\partial}:\theta\bar{\theta}:$ has conformal
dimension of (0,1) and it is similar to the field $\xi$ in the
equation (\ref{T and pesi}). Ignoring the first term, up to a
rescaling $\tilde{\psi}=-2\psi$ the
other terms have the general logarithmic form.\\
Operation of the Virasoro algebra generators $L_n$'s on the rescaled
field $\tilde{\psi}$ yields

\begin{eqnarray}\label{Lnpsi}
L_{-1}\tilde{\psi}=\partial\tilde{\psi},    \hspace{2cm}
L_0\tilde{\psi}&=&\tilde{\psi}+\phi, \nonumber\\
L_1 \tilde{\psi}=\frac{1}{2}\bar{\partial} \tilde{I}, \hspace{2cm}
L_n\tilde{\psi}&=&0 \hspace{0.5cm}for \hspace{0.5cm}n>1.
\end{eqnarray}
Moreover, the OPEs of the field $\psi$ with $\phi$ is explicitly
calculated using the Wick theorem

\begin{eqnarray}\label{sifi}
\phi(z)\psi(w)=&-&\frac{:\theta\bar{\theta}:(w)}{2|z-w|^4}-\frac{\phi(w)}{2|z-w|^2}
+\frac{\bar{\partial}:\theta\bar{\theta}:(w)}{4|z-w|^2(z-w)}+
\frac{\partial:\theta\bar{\theta}:(w)}{4|z-w|^2(\bar{z}-\bar{w})}\nonumber\\&+&\frac{\bar{t}(w)}{2(z-w)^2}
+\frac{t(w)}{2(\bar{z}-\bar{w})^2}-\frac{\partial\bar{t}(w)}{4(z-w)}+\cdot\cdot\cdot.
\end{eqnarray}
In this OPE one can see that the two operators
$t=2:\partial\theta\partial\bar{\theta}\theta\bar{\theta}:$ and
$\bar{t}=2:\bar{\partial}\theta\bar{\partial}\bar{\theta}\theta\bar{\theta}:$
appear which we name them logarithmic energy-momentum tensor. We
can not write a closed algebra for the modes of $\psi$ as we did
in equation (\ref{mode expansion}). Moreover the OPEs of $\psi$
with itself and $:\theta\bar{\theta}:$ are given in the appendix,
these OPEs are useful in calculating the RG equations.

Note that the equations (\ref{sifi}) and (\ref{sisi}) don't
satisfy
the ordinary logarithmic OPEs. \\
It is not difficult to calculate all of the two point correlation
functions including the fields $\psi$ and $\phi$. One can obtain
them by using the wick theorem and the original correlations

\begin{eqnarray}\label{csisi}
\langle\phi(z)\phi(w)\rangle&=&0 ,\\
\langle\psi(z)\phi(w)\rangle&=&\frac{1}{2|z-w|^4},\\
\langle\psi(z)\psi(w)\rangle&=&\frac{1}{2|z-w|^4}\{1+\log|z-w|^2\},\label{csisi2}\\
\langle\phi(z)\bar{\partial}:\theta\bar{\theta}:(w)\rangle&=&0,\\
\langle\psi(z)\bar{\partial}:\theta\bar{\theta}:(w)\rangle&=&\frac{1}{2(z-w)(\bar{z}-\bar{w})^{2}}\label{csisi3}.
\end{eqnarray}
The above correlators are similar to ones which were investigated
in the first section, here we have $h=1$. It is interesting that
the two point functions of the fields with different weights can
be found by using just the OPEs of these fields with
energy-momentum tensor.\\
\newpage
\subsection{W algebra and the Highest Weight Representations of
\\$c=-2$ Model  }\

We conclude this section by relating the above calculations to the
algebraic approach of Gaberdiel and Kausch \cite{g}. Using Zhu's
algebra, one can show that the representation which contains $I$
and $\tilde{I}$ is the only logarithmic one which is generated
from a highest weight state, but the other logarithmic
representations which may be constructed by other methods can not
be found in this way. An example is the representation generated
from $\phi$ and $\tilde{\psi}$ with the conformal weight of one,
and $\xi$ with the weight of zero named $\mathcal{R_{1}}$ (see
\cite{g} for more details).\\ The method of Gaberdiel and Kausch
is based on the $W$ algebra structure of the $c=-2$ model. One may
simply check that all of the following fields have conformal
dimension of three

\begin{eqnarray}\label{wfield}
W^{+}&=&\partial^{2}\theta \partial\theta \nonumber\\
W^{0}&=&\frac{1}{2}(\partial^{2}\theta\partial\bar{\theta}+\partial^{2}\bar{\theta}\partial\theta)\\
W^{-}&=&\partial^{2}\bar{\theta} \partial\bar{\theta}\nonumber.
\end{eqnarray}
The above fields are isospin one fields and have the following
OPEs with themselves \cite{kl}.

\begin{eqnarray}\label{wfield}
W^{i}(z)W^{j}(w)&=&
g^{ij}(\frac{1}{(z-w)^{6}}-3\frac{T(w)}{(z-w)^{4}}-\frac{3}{2}\frac{\partial
T(w)}{(z-w)^{3}}
+\frac{3}{2}\frac{\partial^{2}T(w)}{(z-w)^{2}}\nonumber\\&-&4\frac{T^{2}(w)}{z-w}
+\frac{1}{6}\frac{\partial^{3}T(w)}{z-w}-4\frac{\partial
T^{2}(w)}{z-w})
-5f^{ij}_{k}(\frac{W^{k}(w)}{(z-w)^{3}}\nonumber\\&+&\frac{1}{2}\frac{\partial
W^{k}}{(z-w)^{2}}+
\frac{1}{25}\frac{\partial^{2}W^{k}(w)}{z-w}+\frac{1}{25}\frac{(TW^{k})(w)}{z-w}),
\end{eqnarray}
where $g^{ij}$ is the metric on the isospin one representation,
$g^{+-}=g^{-+}=2$ and $g^{00}=-1$, and  $f^{ij}_{k}$ are the
structure constants of $SL(2)$.\\ By using the above OPEs, one can
write the $W$ algebra which was first investigated by
Zamolodchikov \cite{zamol}. Following Gaberdiel and Kausch
\cite{kg1} we have

\begin{eqnarray}\label{walgebra}
[L_m,W^i_n]&=&(2m-n) W^i_{m+n}\\ \ \
[W_m^i,W_n^j]&=&
g^{ij}(2(m-n)\Lambda_{m+n}+\frac{1}{20}(m-n)(2m^2+2n^2-m n-8)
L_{m+n}
\nonumber \\ &-& \frac{1}{120} m(m^2-1)(m^2-4)\delta_{m+n} ) ) \nonumber \\
&+&f^{ij}_k \left( \frac{5}{14} (2m^2+2n^2-3mn-4)W^k_{m+n} +
\frac{12}{5} V^k_{m+n} \right)
\end{eqnarray}
where $\Lambda = \mathopen:T^2\mathclose: - 3/10\, \partial^2T$
and $V^a = \mathopen:TW^a\mathclose: - 3/14\,
\partial^2W^a$ are quasiprimary normal ordered fields. The above
algebra is a little different from the original Zamolodchikov's
$W$ algebra, the last term multiplied by the factor $f^{ij}_k$ is
different.\\ By using the above $W$ algebra, Gaberdiel and Kausch
found two following relevant nontrivial null vectors

\begin{eqnarray}\label{walgebra}
N^{i}&=&(2L_{-3}W_{-3}^{i}-\frac{4}{3}L_{-2}W_{-4}^{i}+W_{-6}^{i})I,\\
N^{ij}&=&W_{-3}^{i}W_{-3}^{j}I-g^{ij}\left(\frac{8}{9}L_{-2}^{3}+
\frac{19}{36}L_{-3}^{2}+\frac{14}{9}L_{-4}L_{-2}-\frac{16}{9}L_{-6}\right)I\cr
&-&f^{ij}_k\left(-2L_{-2}W_{-4}^{k}+\frac{5}{4}W_{-6}^{k}\right)I.
\end{eqnarray}

Actually, one can simply check that the field $W^{0}$ is a level
two descendent of some fields which will be investigated later. We
use the above null vectors to determine the allowed highest weight
representations of $c=-2$ model. Any correlator involving the
states in a representation of CFT and null vectors must vanish.
One can simplify this condition by stating that the operation of
the zero mode of a null vector on every highest weight state must
vanish. These modes are named Zhu's modes. The zero modes of
(\ref{walgebra}) enforce the following relation for any arbitrary
highest weight field $\varphi$ \cite{g,ng}

\begin{eqnarray}\label{zhu}
L_{0}^{2}(8L_{0}+1)(8L_{0}-3)(L_{0}-1)\varphi=0.
\end{eqnarray}

The above equation implies that $h$ must be $h=0,-1/8,3/8,1$. In
other words, by evaluating the constraints which come from the null
vectors, we find that for irreducible representations, we have the
highest weight representations just for the fields with the
above weights. The representations are named $\mathcal{V}$$_{h}$.\\
The $h=0$ representation is a logarithmic representation but the
others are not, although the fusion products of some of them can
produce logarithmic representations. Let's first investigate
$\mathcal{V}$$_{0}$. This representation comes from
$L_{0}^{2}\varphi=0$ which can be satisfied by $\tilde{I}$ and
$I$ where $L_{0}\tilde{I}=I$ and $L_{0}I=0$. This representation
is the only logarithmic highest weight representation of $c=-2$
model. Although, there are many other logarithmic representations
but they are not highest weight representations. In every
logarithmic theory, we must have a term like $(L_{0}-h)^{n}$ in
the Zhu's algebra for rank $n$ Jordan cells with a logarithmic
highest weight representation to exist. In $c=-2$ model, there
are three other highest weight representations which two of them
are related to the twisted sector and the other is nontwisted. To
complete the classification of the highest weight
representations, one can write the following algebra for the zero
modes of $W$ algebra by using the constraint coming from the zero
modes of the null vectors:

\begin{eqnarray}\label{su2}
[W_{0}^{i},W_{0}^{j}]=\frac{2}{5}(6h-1) f_{k}^{ij}W_{0}^{k}.
\end{eqnarray}

This is similar to the $SU(2)$ algebra, so we can use it to label
the representations. Like the ordinary representations of $SU(2)$,
$j$ and $m$ label the representations. For $c=-2$ there is another
constraint, $W_{0}^{i}W_{0}^{i}=W_{0}^{j}W_{0}^{j}$. So we must
have $j(j+1)=3m^{2}$. With this constraint, one can find that just
representations $j=0,\frac{1}{2}$ exist. By using (\ref{su2}), we
simply find that the representations $\mathcal{V}$$_{0}$ and
$\mathcal{V}$$_{-1/8}$ are related to $j=0$ but the others are
related to $j=1/2$. So, we have two singlet representations and
two doublet representations. These four irreducible highest weight
representations can be multiplied together to produce the well
known representations \cite{g}. The fusion product of the above
representations have the following forms

\begin{eqnarray}\label{fusion}
{\mathcal{V}}_{0}\otimes
{\mathcal{V}}_{h}&=&{\mathcal{V}}_{h}\nonumber\\
{\mathcal{V}}_{1}\otimes
{\mathcal{V}}_{1}={\mathcal{V}}_{0}\hspace{2cm}
{\mathcal{V}}_{0}\otimes
{\mathcal{V}}_{-1/8}&=&{\mathcal{V}}_{3/8}\hspace{2cm}{\mathcal{V}}_{0}\otimes
{\mathcal{V}}_{3/8}={\mathcal{V}}_{-1/8}\nonumber\\
{\mathcal{V}}_{-1/8}\otimes
{\mathcal{V}}_{-1/8}={\mathcal{R}}_{0}\hspace{2cm}{\mathcal{V}}_{-1/8}\otimes
{\mathcal{V}}_{3/8}&=&{\mathcal{R}}_{1}\hspace{2cm}{\mathcal{V}}_{3/8}\otimes
{\mathcal{V}}_{3/8}={\mathcal{R}}_{0}
\end{eqnarray}
where $\mathcal{R}$$_{0}$ and $\mathcal{R}$$_{1}$ are the
representations which were introdused in the previous sections. To
complete this argument, we must compute the behavior of operators
under the $W$ algebra modes. first observe that
$\mathcal{R}$$_{0}$ is a singlet under the $W$ algebra. One can
simply show that $W_{n}^{i}\tilde{I}=0$ for all $n\geq 0$.
Moreover, $\mathcal{R}$$_{1}$ is a doublet under the $W$ algebra
modes and also can be simply shown that
\begin{eqnarray}\label{wmode}
W_{0}^{i}\varphi^{\alpha}=2t_{\beta}^{i\alpha}\varphi^{\beta}\hspace{2cm}
W_{-1}^{i}\theta^{\alpha}=t_{\beta}^{i\alpha}\varphi^{\beta}\hspace{2cm}W_{0}^{i}\theta^{\alpha}=0\nonumber\\
W_{0}^{i}\psi^{\alpha}=2t_{\beta}^{i\alpha}\psi^{\beta}\hspace{2cm}
W_{1}^{i}\varphi^{\alpha}=t_{\beta}^{i\alpha}\theta^{\beta}\hspace{2cm}
\end{eqnarray}
where $t_{\beta}^{i\alpha}$ is the spin $1/2$ representation of
$SU(2)$ where just the elements $t_{\pm}^{0\pm}=\pm1/2$ and
$t_{\pm}^{\pm\mp}=1$ are not zero.

The other important property is that one can calculate the other
fusion products and prove that the  four representations
$\mathcal{V}$$_{h}$ with allowed weights and $\mathcal{R}$$_{0}$ and
$\mathcal{R}$$_{1}$ are closed under fusion. It means that we have a
rational logarithmic conformal field theory. Finally the
representation $\mathcal{R}$ is a local representation which is a
direct sum of $\mathcal{R}$$_{0}$, $\mathcal{R}$$_{1}$ and
$\mathcal{V}$$_{h}$ with some subtractions. For $\mathcal{R}$ to be
local, we need to have $h-\bar{h}\in$$\mathcal{Z}$ and
$S\phi=(L_{0}^{n}-\bar{L}_{0}^{n})\phi=0$ where $\phi$ is a operator
in the local theory \cite{g}. The fields in $\mathcal{R}$ satisfy
these properties and it is a local triplet representation for $c=-2$
model.

\section{Logarithmic Energy-Momentum Tensor and Logarithmic \\Sugawara
Construction}\ \setcounter{equation}{0}\label{Sugawara}\

The logarithmic partner of the energy-momentum tensor $t$ can be
constructed using the fields $T$ and $:\theta\bar{\theta}:$,
namely $:T \theta\bar{\theta}:$ . Despite the fact that $t$ has
been known to exist for a while, its OPEs were not well known.
The OPE of $t$ with $T$ takes on the form:

\begin{eqnarray}\label{tT}
\hspace{1cm}
T(z)t(w)=-\frac{:\theta\bar{\theta}:(w)}{(z-w)^4}+\frac{\partial:\theta\bar{\theta}:(w)}{(z-w)^3}
+\frac{2t(w)}{(z-w)^2}-\frac{T(w)}{2(z-w)^2} +\frac{\partial
t(w)}{z-w}+\cdot\cdot\cdot.
\end{eqnarray}
Let us rescale $\tilde{t}=-2t$ in order to find familiar
equations. We can see that the Virasoro operators on $\tilde{t}$
take the form:

\begin{eqnarray}\label{Lnt}
L_{-2}\tilde{I}&=&\tilde{t},    \hspace{3cm}
L_{-2}I=T \nonumber\\
L_{-1}\tilde{t}&=&\partial\tilde{t},    \hspace{3cm}
L_0\tilde{t}=2\tilde{t}+T \nonumber\\
L_1 \tilde{t}&=&\partial \tilde{I}, \hspace{3cm}L_2
\tilde{t}=-\tilde{I}\nonumber\\
 L_n\tilde{t}&=&0
\hspace{1.5cm}for \hspace{0.5cm}n>2.
\end{eqnarray}
Let us now calculate the finite transformation of the operator $t$
using the OPE (\ref{tT}):
\begin{eqnarray}\label{finitet}
t(z)&=&\left(\frac{\partial w}{\partial
z}\right)^{2}t(w)-\frac{1}{2}\log\left(\frac{\partial w}{\partial
z}\right)\left((\frac{\partial w}{\partial
z})^{2}T(w)-\frac{1}{6}\{w,z\}\right)\nonumber\\&+&\frac{1}{2}\frac{\partial^{2}w}{\partial^{2}z}
\left(\partial:\theta\bar{\theta}:(w)-\frac{\frac{\partial^{2}w}{\partial^{2}z}}{2(\frac{\partial
w}{\partial z})^{2}}\right)-\frac{1}{6}
\{w,z\}\left(:\theta\bar{\theta}:(w)-\frac{1}{2}\log(\frac{\partial
w}{\partial z})\right)
\end{eqnarray}
where the $\{w,z\}$ is the schwarzian derivative defined by
\begin{eqnarray}\label{schwarzian}
\{w,z\}=\frac{\partial_{z}^{3}w}{\partial_{z}^{2}w}-\frac{3}{2}\left(\frac{\partial_{z}^{2}w}{\partial_{z}w}\right)^{2}.
\end{eqnarray}
To find equation (\ref{finitet}) we used the general conformal
transformation of the energy-momentum tensor
\begin{eqnarray}\label{finiteT}
T(z)=\left(\frac{\partial w}{\partial
z}\right)^{2}T(w)-\frac{1}{6}\{w,z\}.
\end{eqnarray}
Nevertheless, here the two point functions including the fields
$t$ and $T$ have their standard logarithmic form:
\begin{eqnarray}\label{cT1T}
\langle T(z)T(w)\rangle&=&0,\\
\langle t(z)T(w)\rangle&=&\frac{1}{(z-w)^4},\\
\langle
t(z)t(w)\rangle\hspace{.2cm}&=&\frac{1}{(z-w)^4}\{1+\log|z-w|^2\}.
\end{eqnarray}
The correlation functions of $T$ with the other operators vanish
but the one for $t$ have nonzero values. For example, the two
point function of $t$ with $:\theta\bar{\theta}:$ is :

\begin{eqnarray}\label{ttetta}
\langle t(z):\theta\bar{\theta}:(w)\rangle=\frac{1}{2(z-w)^2}.
\end{eqnarray}
The OPEs of $t$ with the fields $:\theta\bar{\theta}:$, $\psi$
and $\phi$ are given in the appendix.

The other important OPE is $t$ with itself. as we can see from the
above nontrivial correlation functions it does not have a simple
form:
\begin{eqnarray}\label{tt}
t(z)t(w)&=&\frac{I}{4(z-w)^4}\{1+2\log|z-w|^2+\log^2|z-w|^2\}\nonumber\\
&-&\frac{1}{(z-w)^4}\{1+\log|z-w|^2\}:\theta\bar{\theta}:(w)
+\frac{1}{2(z-w)^2}\log|z-w|^2
t(w)\nonumber\\&-&\frac{1}{2(z-w)^2}\log|z-w|^2\{1+\log|z-w|^2\}T(w)+\cdot\cdot\cdot.
\end{eqnarray}

The existence of logarithms do not allow derivation of a closed
algebra for the modes of $t$. One can write a different
logarithmic energy-momentum tensor which is related to the
Logarithmic Sugawara construction. Logarithmic Sugawara
construction is defined with a little defect. Given the primary
pre-logarithmic currents $J^{i}$ then we can define the
energy-momentum tensors as :

\begin{eqnarray}\label{logarithmic sugawara}
J^{i}(z)J^{i}(0)\sim \cdot\cdot\cdot+(T \log z+\hat{t}
)+\cdot\cdot\cdot.
\end{eqnarray}

For using the above construction for $c=-2$ model we need some
currents. The action (\ref{s}) is invariant under $SL(2)$
transformations on the fields $\theta^{\alpha}$ which their
infinitesimal transformation is

\begin{eqnarray}\label{sl2}
\delta\theta^{\alpha}=-i\Lambda_{i}(z,\bar{z})t_{\beta}^{i\alpha}\theta^{\beta}.
\end{eqnarray}
So by using Noether's theorem and some rescaling we can write the
following currents

\begin{eqnarray}\label{currents}
J^{+}&=&\frac{4}{3}\theta\partial\theta\nonumber,\\
J^{0}&=&\frac{2}{3}(\theta\partial\bar{\theta}+\bar{\theta}\partial\theta),\\
J^{-}&=&\frac{4}{3}\bar{\theta}\partial\bar{\theta}.\nonumber
\end{eqnarray}

One can simply prove that all of these operators are primary
fields with conformal dimensions equal to one. In fact these
operators are related to the fields of $\mathcal{R}$
representation. Like the previous sections by using fundamental
correlations one can find the following OPE for the currents

\begin{eqnarray}\label{opecurrents}
J^{i}(z)J^{j}(0)=g^{ij}\left(\frac{\log
zI+\tilde{I}+I}{z^2}+\frac{\partial\tilde{I}}{2z}\right)+\frac{f^{ij}_{k}J^{k}}{z}+\cdot\cdot\cdot,
\end{eqnarray}
which is similar to affine Lie algebra. By looking at
equation(\ref{logarithmic sugawara}), one can guess that the
$\hat{t}$ must have the following form

\begin{eqnarray}\label{t suga}
\hat{t}=-2t-\frac{3}{2}\partial^{2}\tilde{I}-5T.
\end{eqnarray}
$\hat{t}$ is a descendent field of $I$ and $\tilde{I}$, and may be
written as follows:

\begin{eqnarray}\label{descendant t suga}
\hat{t}=(L_{-2}-\frac{5}{2}L_{-1}^{2})\tilde{I}-4L_{-2}I.
\end{eqnarray}
It is not difficult to calculate the two point functions of this
field with $T$ and itself. These are similar to the two point
functions of $t$ case whereas its OPEs are different.

\setcounter{equation}{0}
\section{Kac Table in $c=-2$ Theory }\

Usually the operator content of minimal models
$\mathcal{M}$$(p,p')$ with $p>p'$ is the $\phi_{r,s}$ such that
$0<r<p'$, $0<s<p$. For the $c=-2$ model the Kac table is the case
with $(p,p')=(2,1)$ and it is trivial. The solution is to extend
the Kac table to find a nontrivial operator content. The Kac
formula of conformal dimension  $h_{r,s}$ for minimal model
$\mathcal{M}$$(p,p')$ is given as

\begin{eqnarray}\label{Kac}
h_{r,s}=\frac{(pr-p's)^{2}-(p-p')^{2}}{4pp'}.
\end{eqnarray}
The formula for $h_{r,s}$ has an important symmetry
$h_{r,s}=h_{p-r,q-s}$. For $c=-2$ model we have the following
formulas
\begin{eqnarray}\label{Kac2}
h_{r,s}=\frac{(2r-s)^{2}-1}{8},\hspace{2cm}h_{r,s}=h_{1-r,2-s}=h_{r-1,s-2}
\end{eqnarray}
\begin{table}[htb]
\begin{center}
\caption{Extended Kac Table for c=-2 Model  }\

\begin{tabular}{c|cccccccccccc}
\hline
&\multicolumn{12}{c}{$s$}\\
\cline{2-13}
$r$&$1$&$2$&$3$&$4$&$5$&$6$&$7$&$8$&$9$&$10$&$11$&$12$\\
\hline\hline
$1$&$0$&$-1/8$&$0$&$3/8$&$1$&$15/8$&$3$&$35/8$&$6$&$63/8$&$10$&$\cdots$\\
$2$&$1$&$3/8$&$0$&$-1/8$&$0$&$3/8$&$1$&$15/8$&$3$&$35/8$&$6$&$\cdots$\\
$3$&$3$&$15/8$&$1$&$3/8$&$0$&$-1/8$&$0$&$3/8$&$1$&$15/8$&$3$&$\cdots$\\
$4$&$6$&$35/8$&$3$&$15/8$&$1$&$3/8$&$0$&$-1/8$&$0$&$3/8$&$1$&$\cdots$\\
$5$&$10$&$63/8$&$6$&$35/8$&$3$&$15/8$&$1$&$3/8$&$0$&$-1/8$&$0$&$\cdots$\\
$6$&$15$&$99/8$&$10$&$63/8$&$6$&$35/8$&$3$&$15/8$&$1$&$3/8$&${0}$&$\cdots$\\
$7$&$\dots$&${}$&${}$&${}$&${}$&${}$&${}$&${}$&${}$&${}$&${}$&$\cdots$\\
\hline
\end{tabular}
\end{center}
\end{table}
It is interesting to relate the primary fields which we
investigated in the previous sections to the the fields in the Kac
table. The simplest one is the field $\phi_{1,1}$ which is the
identity operator. The field $\tilde{I}$ is equivalent to
$\phi_{1,3}$ which can be simply checked that this field has a
null vector at level three. It is a property of Kac table fields
that the field $\phi_{r,s}$ has null vector descendants at levels
$rs$ and $(1-r)(2-s)$. So the field $\tilde{I}$ has null vector
at level three. The other field is $\phi_{2,1}$ which is related
to $\varphi^{\alpha}$ operators. Its logarithmic partner
$\psi^{\alpha}$, as Flohr showed \cite{Flohr}, is related to
$\phi_{1,5}$. The field $\phi_{1,2}$ is the twist operator $\mu$
with the null vector at level two.  The doublet of twist fields
$\sigma_{\alpha}=(\theta_{\alpha})_{-\frac{1}{2}}\mu$ of dimension
$\frac{3}{8}$ is equivalent to $\phi_{2,2}$ and $\phi_{1,4}$ in
the Kac table. The other interesting fields are the $W^{i}$ fields
which are related to the $\phi_{3,1}$, the logarithmic partner of
this field is $\phi_{1,7}$. So one can find an algorithm to
relate the primary fields to the logarithmic partners, the
logarithmic partner of $\phi_{r,1}$ is $\phi_{1,2r+1}$. One must
notice that $\phi_{r,s}=\phi_{r+1,s+2}$ and so the fields
$\phi_{2,2}$ and $\phi_{3,4}$ are equivalent. This important
result states that just the two lines on the boundary of grid are
important and produce new fields. Another example is the fields
equivalent to the $\phi_{4,1}$ with null vector at level four and
the conformal weight of three
\begin{eqnarray}\label{H}
H^{\frac{3}{2}}&=&\partial^{3}\theta\partial^{2}\theta\partial\theta\nonumber\\
H^{\frac{1}{2}}&=&\frac{1}{3}(\partial^{3}\theta\partial^{2}\theta\partial\bar{\theta}+
\partial^{3}\theta\partial^{2}\bar{\theta}\partial\theta+\partial^{3}\bar{\theta}\partial^{2}\theta\partial\theta)\nonumber\\
H^{\frac{-1}{2}}&=&\frac{1}{3}(\partial^{3}\theta\partial^{2}\bar{\theta}\partial\bar{\theta}+
\partial^{3}\bar{\theta}\partial^{2}\bar{\theta}\partial\theta+\partial^{3}\bar{\theta}\partial^{2}\theta\partial\bar{\theta})\nonumber\\
H^{\frac{-3}{2}}&=&\partial^{3}\bar{\theta}\partial^{2}\bar{\theta}\partial\bar{\theta}.
\end{eqnarray}

Generally, one can check that $\phi_{r,1}$ generates Isospin
$\frac{r-1}{2}$ fields,. So in level $(r,1)$ we have $r$ fields
with conformal dimension of $\frac{r^{2}-r}{2}$. For example in
level $(5,1)$ we have the following primary fields
\begin{eqnarray}\label{M}
M^{2}&=&\partial^{4}\theta\partial^{3}\theta\partial^{2}\theta\partial\theta\nonumber\\
M^{1}&=&\frac{1}{4}(\partial^{4}\theta\partial^{3}\theta\partial^{2}\theta\partial\bar{\theta}+
\partial^{4}\partial^{3}\theta\partial^{2}\bar{\theta}\partial\theta+
\partial^{4}\theta\partial^{3}\bar{\theta}\partial^{2}\theta\partial\theta+
\partial^{4}\bar{\theta}\partial^{3}\theta\partial^{2}\theta\partial\theta)\nonumber\\
M^{0}&=&\frac{1}{6}(\partial^{4}\theta\partial^{3}\theta\partial^{2}\bar{\theta}\partial\bar{\theta}+
\partial^{4}\theta\partial^{3}\bar{\theta}\partial^{2}\theta\partial\bar{\theta}+
\partial^{4}\bar{\theta}\partial^{3}\theta\partial^{2}\theta\partial\bar{\theta}+
\partial^{4}\theta\partial^{3}\bar{\theta}\partial^{2}\bar{\theta}\partial\theta\nonumber\\&+&
\partial^{4}\bar{\theta}\partial^{3}\theta\partial^{2}\bar{\theta}\partial\theta+
\partial^{4}\bar{\theta}\partial^{3}\bar{\theta}\partial^{2}\theta\partial\theta
)\nonumber\\
M^{-1}&=&\frac{1}{4}(\partial^{4}\bar{\theta}\partial^{3}\bar{\theta}\partial^{2}\bar{\theta}\partial\theta+
\partial^{4}\bar{\theta}\partial^{3}\bar{\theta}\partial^{2}\theta\partial\bar{\theta}+
\partial^{4}\bar{\theta}\partial^{3}\theta\partial^{2}\bar{\theta}\partial\bar{\theta}+
\partial^{4}\theta\partial^{3}\bar{\theta}\partial^{2}\bar{\theta}\partial\bar{\theta})
\nonumber\\
M^{-2}&=&\partial^{4}\bar{\theta}\partial^{3}\bar{\theta}\partial^{2}\bar{\theta}\partial\bar{\theta}.
\end{eqnarray}
Can be checked that all of the above fields have null vectors at
level five. The higher primary conformal fields can be found in
the same way. \\One can also find the fermionic representation of
other primary fields in the Kac table. The column $(r,2)$ is
related to the twisted sector. Suppose in (\ref{te}) we take
$n=m-\tau$ with integer $m$ and half integer $\tau$; then the
fields $\nu_{\alpha}=(\theta_{\alpha})_{-\tau}\mu$ have conformal
weight $h_{\tau}=\frac{\tau(\tau-1)}{2}$. They form the second
column in the Kac table, for example $\phi_{3,2}$ and
$\phi_{1,6}$ are related to $\tau=\frac{5}{2}$. Generally, the
fields $\phi_{r,2}$ and $\phi_{1,2r}$ have the same conformal
weights.

\subsection{Hierarchy of W algebras}\

In section (\ref{Logarithm}) we saw that there exists a W(2,3,3,3)
algebra with the primary operators $W^{i}$ and conformal
dimension of three. These fields are descendants of the $J^{i}$
currents:

\begin{eqnarray}\label{null w}
(L_{-2}-\frac{1}{2}L_{-1}^{2})J^{i}=2W^{i}.
\end{eqnarray}
Because of the absence of zero modes. the two point correlation
functions of all these fields with themselves vanish. Notice that
the existence of zero modes is not sufficient for nontrivial
correlation functions, the best examples are the logarithmic
partners of $W^i$ fields which are produced by insertion of
$\tilde{I}$ aside them. Their two point functions have the
following forms
\begin{eqnarray}\label{two point w}
\langle W^{i}(z)\tilde{W}^{j}(w)\rangle&=&g^{ij}\frac{2}{(z-w)^{6}},\\
\langle\tilde{W}^{i}(z)\tilde{W}^{j}(w)\rangle&=&g^{ij}\frac{2}{(z-w)^{6}}(1+\log
z).
\end{eqnarray}
The fields $\tilde{W}^{i}$ have the following OPE with
energy-momentum tensor
\begin{eqnarray}\label{ope T w}
T(z)\tilde{W}^{i}(w)=\frac{-3J^{i}}{2(z-w)^{4}}+\frac{K^{i}}{2(z-w)^{3}}+\frac{3\tilde{W}+W}{(z-w)^{2}}+\frac{\partial\tilde{W}}{z-w},
\end{eqnarray}
where $K^{i}$s are non-primary fields with the following explicit
forms :

\begin{eqnarray}\label{K}
K^{+}=\theta\partial^{2}\theta,\hspace{1.5cm}
K^{0}=\theta\partial^{2}\bar{\theta}+\bar{\theta}\partial^{2}\theta,\hspace{1.5cm}
K^{-}=\bar{\theta}\partial^{2}\bar{\theta}.
\end{eqnarray}
Whereas the operator $W^{0}$ has a null vector at level three,
the operators $W^{+}\theta$, $W^{-}\bar{\theta}$ and
$W^{0}\bar{\theta^{\alpha}}$ do not have a null vector at level
three. Operating the following level three operator on these
fields, one can find the $\frac{1}{4}H^{i}$ fields
\begin{eqnarray}\label{level three null}
(L_{-3}-\frac{2}{5}L_{-1}L_{-2}+\frac{1}{20}L_{-1}^{3}).
\end{eqnarray}
This is similar to equation (\ref{null w}), in fact the same
relations exist for higher level primary fields. For example, by
operating the level four null operator on the
$H^{\frac{3}{2}}\theta$, one can obtain the operators proportional
to the $M^{2}$.

The above argument comes from the relation $h_{r,1}+r=h_{r+1,1}$
in which $r$ is the level of the null vector of $\phi_{r,1}$. It
does not indicate that $W(2,6,6,6,6)$, $W(2,6)$,
$W(2,10,10,10,10,10)$ or $W(2,10)$ algebras in $c=-2$ model exist.
 \cite{kw,behhh} show that $W(2,10)$ algebra is consistent with
$c=-2$ whereas the $W(2,6)$ is inconsistent. In \cite{eh} the
authors used the $W(2,10)$ algebra and found some of the results
which we have established in section (\ref{Logarithm}). Our
calculations are consistent with the existence of some other $W$
algebras. However this is not quite clear yet and work in this
direction is under progress.\\
These $W$ algebras are not the only algebras which appeare in the
$c=-2$ model. One of the other closed complex algebras is the
(\ref{mode expansion}) algebra which mixes the holomorphic and
antiholomorphic parts of the Virasoro algebra; a property which
can not be seen in the ordinary $W$ algebras.\\ All of these
expressions show that the action (\ref{s}) is a good
representation for $c=-2$ conformal field theory. This action has
suitable representations for all of the fields which appear in the
Kac table.

In the next sections we will use some of the fields introduced
before to perturb the $c=-2$ theory. In this way, we will
investigate some off-critical models which are not conformally
invariant but their contents are similar to the ordinary $c=-2$
model.

\setcounter{equation}{0}
\section{Perturbation of the $c=-2$ Theory }\label{Perturb}\

Conformal field theories describe the behavior of a system at its
critical point. Also in two dimensions CFT's are integrable since
the conformal algebra is infinite dimensional. Therefore it is
natural to expect that perturbation of a CFT by an operator may
lead to an integrable model in two dimensions \cite{zamolo}.
However not all perturbations may lead to integrable models. The
perturbing operator has to be chosen carefully so that an infinite
number of currents remain conserved, therefore the structure of
the CFT becomes important. The case for unitary models with
central charge $c=1-\frac{6}{p(p+1)} , p=3,4,5,...$, and
perturbing fields being $\phi_{1,2},\phi_{2,1},\phi_{1,3}$ was
analyzed in \cite{zamolo}. The usefulness of this approach lies in
the fact that one may use the structure of CFT to investigate the
integrable models. In fact, using this device, Zamolodchikov
solved the Ising model in two dimensions in presence of magnetic
field \cite{zamol,delfino}. In this section, we investigate
theories which are obtain from perturbation of $c=-2$ theory by
fields such as $:\theta\bar{\theta}:$ and
$:\partial\theta\bar{\partial}\theta:$ or powers of the
energy-momentum tensor $T_{2n}$ . Moreover we briefly discuss the
integrability of these theories. We just focus on the conserved
currents; the exact study needs the consistency of $S$ matrix .

The perturbed action is obtained from the critical action $S^{*}$
by the addition of the operator $\Phi(z,\bar{z})$,
\begin{eqnarray}\label{tperturbed}
S=S^* + \alpha\int d^2z\Phi(z,\bar{z}).
\end{eqnarray}

For clarification, let us investigate the conservation of
energy-momentum tensor in a typical theory perturbed with an
arbitrary scaling field $\Phi$ with conformal dimension of $h$.\\
The correlation functions of a particular operator $J(z,\bar{z})$
are given by the following equation:

\begin{eqnarray}\label{cJ}
\langle J(z,\bar{z}) \cdot\cdot\cdot\rangle=\langle J(z,\bar{z})
\cdot\cdot\cdot\rangle_{s^*}+\alpha\int d^2z_1\langle
J(z,\bar{z})\Phi(z_1,\bar{z}_1)
\cdot\cdot\cdot\rangle_{s^*}+{\mathcal{O}}(\alpha^2).
\end{eqnarray}
Generally, the $\bar{z}$ dependence of the finite case of this
integral emerges from the singularities in the neighborhood of
$z_1$. So, in this limit i.e. $z\rightarrow z_1 $, we can use the
following OPE in the above integral

\begin{eqnarray}\label{Jfi}
J(z,\bar{z})\Phi(z_1,\bar{z}_1)=\sum_i
\frac{a_i}{|z-z_1|^{\Delta+\Delta_J-\Delta_i}}\phi_i(z_1,\bar{z}_1),\hspace{4cm}
\end{eqnarray}
where $\Delta=2h$ and $\Delta_J$ and $\Delta_i$ are the scaling
dimension of the fields $J$ and $\phi_i$ respectively. These
singularities are integrable for $\Delta+\Delta_J-\Delta_i<2$.
Since in a unitary theory all dimensions are positive, only a
finite number of operators $\phi_i$ contribute in the correlation
expansion up to the first order in $\alpha$.\\ In particular for
the energy-momentum tensor, the OPE is:

\begin{eqnarray}\label{Tfii}
T(z)\Phi(z_1,\bar{z}_1)=\frac{h}{(z-z_1)^2}\Phi(z_1,\bar{z}_1)
+\frac{1}{z-z_1}\partial_1\Phi(z_1,\bar{z}_1),\hspace{4cm}
\end{eqnarray}
where $\partial_1$ denotes $\partial_{z_1}$.\\
Using the equation (\ref{cJ}) and  regularizing the second term by
cutting out a small section $|z-z_1|^2\leq a^2$, where $a$ is a
microscopic length scale, one can immediately read off the
conservation law for the energy-momentum tensor as;
$\bar{\partial}T+\partial U=0$, where
\begin{eqnarray}\label{U}
U=\pi\alpha(h-1)\Phi(z,\bar{z}).
\end{eqnarray}
Let us now consider particular examples of $\Phi(z,\bar{z})$ for
the $c=-2$ model.

\subsection{Perturbation with $:\theta\bar{\theta}:$ }\

The action of the off-critical, massive theory can be obtained by
perturbation of $c=-2$ theory $S^*$, with the logarithmic partner of
the identity $:\theta\bar{\theta}:$,

\begin{eqnarray}\label{action2}
S=S^* + \frac{m^2}{4}\int:\theta\bar{\theta}:
\end{eqnarray}
The correlation functions can be obtained from the equation of
motion
\begin{eqnarray}\label{tt1}
\langle\theta(z)\bar{\theta}(w)\rangle&=&K_0(m|z-w|),
\hspace{1.2cm}
\langle\theta(z)\theta(w)\rangle=\langle\bar{\theta}(z)\bar{\theta}(w)\rangle=0\\
\langle\partial\theta(z)\partial\bar{\theta}(0)\rangle&=&
-\frac{m^2}{4}\frac{z}{\bar{z}}\{2K''_0(m|z|)-K_0(m|z|)\},
\end{eqnarray}
where the function $K_0$, is the modified Bessel function. The
massless limit of these correlations is exactly what was obtained
before in (\ref{tete}) and (\ref{dtedte}) which contain the zero
modes.
 To investigate the integrability of a theory, we need to look for
conserved currents which can be obtained from the OPE of the
perturbing field ( in this case $:\theta\bar{\theta}:$) and the
quantity under consideration.

 Using the equations of motion, we can show that a class of
operators;
$\tilde{T}_{2n}(z)=2^n:\partial^n\theta\partial^n\bar{\theta}:(z)$
for n=1,2,...,$\infty$  satisfy the continuity equation. To show
this, we look at the OPE of $\tilde{T}_{2n}$s with the perturbing
field $:\theta\bar{\theta}:$

\begin{eqnarray}\label{T2ntete}
\tilde{T}_{2n}(z,\bar{z}):\theta\bar{\theta}:(w,\bar{w})&=&
-\frac{2^{n-1}\{(n-1)!\}^2}{(z-w)^{2n}}+\cdot\cdot\cdot
+\frac{1}{z-w}\partial \tilde{T}_{2(n-1)}(z,\bar{z})+
\cdot\cdot\cdot,
\end{eqnarray}
where the dots denote less singular terms. If we apply the above
method, the following continuity equation can be obtained for
$T_{2n}$:

\begin{eqnarray}\label{conT2n}
\bar{\partial}\tilde{T}_{2n}(z,\bar{z})=\frac{m^2}{4}\partial
\tilde{T}_{2(n-1)}(z,\bar{z}) \hspace{2cm}
n=1,2,\cdot\cdot\cdot,\infty.
\end{eqnarray}
Since there are an infinite number of conserved quantities, this
theory is integrable. In this theory there are many other
conserved quantities which are more familiar in the context of
Virasoro algebra such as $T_{2n}=L_{-2}^{n}I$ for
$n=1,2,3,...,\infty$. With the exception of $T_2$, the first
three conserved quantities have the following explicit shapes in
terms of fundamental fields:

\begin{eqnarray}\label{conservedT2n}
T_{4}&=&\partial^{3}\theta\partial\bar{\theta}+\partial\theta\partial^{3}\bar{\theta}\\
T_{6}&=&\frac{1}{4}\left(\partial^{5}\theta\partial\bar{\theta}+\partial\theta\partial^{5}\bar{\theta}\right)+\frac{1}{16}\tilde{T}_{6}\\
T_{8}&=&\frac{1}{4}\left(\partial^{5}\theta\partial^{3}\bar{\theta}+\partial^{3}\theta\partial^{5}\bar{\theta}\right)+
\frac{1}{24}\left(\partial^{7}\theta\partial\bar{\theta}+\partial\theta\partial^{7}\bar{\theta}\right).
\end{eqnarray}
Of course one can simply check that all of the fields
$\partial^{n}\theta\partial^{m}\bar{\theta}$ are conserved. The
most important examples are the $W^{i}$ fields which are related
to the $W$ algebras investigated before.

\subsection{Perturbation with $\tilde{T}_{2n}$}

Perturbing the action of $c=-2$ theory with the generalized
energy-momentum tensor results in:

\begin{eqnarray}\label{action3}
S=S^* + \frac{\alpha}{\pi}\int \tilde{T}_{2n}\hspace{2cm}
n=1,2,\cdot\cdot\cdot,\infty.
\end{eqnarray}

This theory is integrable only for $n=1$, since infinite number of
$\tilde{T}_{2n}$s are conserved.To investigate the integrability
of the theory, according to the mentioned prescription in the last
subsection, we need the OPE of $\tilde{T}_{2n}$ with $T\equiv T_2$

\begin{eqnarray}\label{T2nT}
\tilde{T}_{2n}(z,\bar{z})T(w,\bar{w})=-\frac{(n!)^2}{(z-w)^{2n+2}}+\cdot\cdot\cdot+
\frac{1}{z-w}\partial \tilde{T}_{2n}(z,\bar{z})+ \cdot\cdot\cdot.
\end{eqnarray}
The continuity equation becomes
\begin{eqnarray}\label{T2nT}
\bar{\partial}\tilde{T}_{2n}(z,\bar{z})=\alpha \partial
\tilde{T}_{2n}(z,\bar{z}) \hspace{2cm}
n=1,2,\cdot\cdot\cdot,\infty.
\end{eqnarray}
This result was predictable, since any conformal field theory
perturbed with energy-momentum tensor has an infinite number of
conserved quantities.\\ Application of the equation of motion for
$n=1$ yields the correlation functions as:

\begin{eqnarray}\label{caction3}
\langle\theta(z,\bar{z})\theta(w,\bar{w})\rangle&=&\langle\bar{\theta}(z,\bar{z})\bar{\theta}(w,\bar{w})\rangle=0,
\nonumber \\
\langle\theta(z,\bar{z})\bar{\theta}(w,\bar{w})\rangle&=&-\log(\bar{z}-\bar{w}).
\end{eqnarray}
Note that these correlations depend only on the anti-holomorphic
terms.

\subsection{Perturbation with $:\partial\theta\bar{\partial}\theta:$ }\

The action of $c=-2$ theory is constructed using a primary field
$\varphi^{\alpha\bar{\alpha}}$ with $\alpha=\bar{\alpha}$ with
conformal dimension of (1,1) which is a member of the
$\mathcal{R}$ representation. It is interesting to add the other
two $\varphi^{\alpha\bar{\alpha}}$s with $\alpha=\bar{\alpha}$ to
the action and see what happens. The action is

\begin{eqnarray}\label{action4}
S=S^* + \frac{\alpha}{\pi}\int:\partial\theta\bar{\partial}\theta:.
\end{eqnarray}
The OPE of perturbing field with the energy-momentum tensor is:

\begin{eqnarray}\label{Tdtedte}
T(z):\partial\theta\bar{\partial}\theta:(w,\bar{w})=
\frac{1}{(z-w)^2}:\partial\theta\bar{\partial}\theta:(z,\bar{z})+\cdot\cdot\cdot.
\end{eqnarray}
To obtain this result, we utilized the equation of motion i.e.
$\partial\bar{\partial}\theta(z,\bar{z})=0$. Moreover, the
continuity equation is $\bar{\partial}T(z)=0$. This means that the
trace of the energy-momentum tensor is zero i.e. the perturbed
theory is still conformal. One can also obtain this result
explicitly by calculating the elements of the energy-momentum
tensor as bellow:

\begin{eqnarray}\label{Tdtedte}
T^{z\bar{z}}=T^{\bar{z}z}=0 ,\hspace{1cm}
T(z)=2:\partial\theta\partial\bar{\theta}:(z),\hspace{1cm}
\bar{T}(\bar{z})=2:\bar{\partial}\theta\bar{\partial}\bar{\theta}:(\bar{z}).
\end{eqnarray}
The correlation functions in the presence of the perturbing term
do not change and the theory remains conformal. All of these
results hold in a theory perturbed by
$:\partial\bar{\theta}\bar{\partial}\bar{\theta}:$ as well.
\setcounter{equation}{0}
\section{Zamolodchikov's c-Theorem and $c=-2$ Theory}\

Zamolodchikov's c-theorem concerns the behavior of a two
dimensional conformal field theory under the renormalization group
(RG) flow. This theorem is correct just for unitary,
renormalizable quantum field theories, and implies that there is a
function $c(g)$ of the coupling constants $g$ which decreases
monotonically under the RG flow \cite{zamol2}. This function has
constant values only at fixed points, the fixed points being
conformally invariant, and at these points Zamalodchikov's
function takes on the value of the central charge of the
corresponding CFT.

Suppose we have an integrable theory with an infinite number  of
conserved quantities:

\begin{eqnarray}\label{conserv}
\bar{\partial}T_{2n}+\partial U_{2(n-1)}=0,
\end{eqnarray}
for any conserved current $T_{2n}$ there is a function $c_{2n}(g)$
which decreases under the RG flow and takes it's constant value at
the fixed point, i.e. the generalized central charge. Consider the
following correlation functions:

\begin{eqnarray}\label{GFH}
\langle T_{2n}(z,\bar{z})T_{2n}(0,0)\rangle&=&\frac{F(|z\bar{z}|)}{z^{4n}},\nonumber\\
\langle T_{2n}(z,\bar{z})
U_{2(n-1)}(0,0)\rangle&=&\frac{G(|z\bar{z}|)}{z^{4n-1}\bar{z}},\\
\langle
U_{2(n-1)}(z,\bar{z})U_{2(n-1)}(0,0)\rangle&=&\frac{H(|z\bar{z}|)}{z^{2(2n-1)\bar{z}^2}}.\nonumber
\end{eqnarray}
Applying the conservation law (\ref{conserv}), one obtains:

\begin{eqnarray}\label{dc2n}
|x|\frac{dc_{2n}}{d|x|}=-8(2n-1)(4n-1)H
\end{eqnarray}
where
\begin{eqnarray}\label{c2n}
c_{2n}(g)=2\{F-2(2n-1)G-(4n-1)H\}.
\end{eqnarray}
Since for a unitary field theory we have $H\geq0$, one concludes
that $\frac{dc_{2n}}{d|x|}\leq0$.\\ By integrating eq.(\ref{dc2n})
we get:

\begin{eqnarray}\label{del2n}
\Delta c_{2n}=\frac{1}{\pi}4(2n-1)(4n-1)\int d^2zz^{4n-3}
\bar{z}\langle U_{2(n-1)}(z,\bar{z})U_{2(n-1)}(0,0)\rangle,
\end{eqnarray}
where $c \equiv c_2$ is the central charge of the corresponding
CFT. This relation is true for any theory irrespective of being
unitary or not.

Since $c=-2$ theory is nonunitary, Zamolodchikov's c-theorem could
not be applied for it. Rather a function does exist but it may not
be monotonically decreasing. But as mentioned at the end of the
last section, since eq.(\ref{del2n}) is usable for both unitary
and nonunitary theories, we can apply it for $c=-2$ theory. In
this way, we utilize the perturbed massive theory (\ref{action2})
as introduced before in section (\ref{Perturb}). Using both
eq.(\ref{del2n}) for $n=1$ and correlation functions (\ref{tt1}),
the central charge of the theory can be exactly obtained equal to
$-2$, whereas according to (\ref{cT1T}) the central charge is
incorrectly zero. This disagreement emerges from the important
role of the zero modes in $c=-2$ theory, because the repeat of the
calculation of two point function (\ref{cT1T}) containing the
zero modes yields :

\begin{eqnarray}\label{T1T12}
\langle T(z)T(w)\bar{\xi}\xi\rangle=- \frac{1}{(z-w)^4},
\end{eqnarray}
giving $c=-2$.\\ Now, if we repeat the procedure for another
conserved quantity $\tilde{T}_{2n}$, we obtain a generalized
central charge $c_{2n}$ for each $n$ as mentioned in \cite{cl}:

\begin{eqnarray}\label{c-2n}
c_{2n}=-2^{2n-1}\{(2n-1)!\}^2.
\end{eqnarray}
In this case, one should again take zero modes into account too.
The correlation function is explicitly obtained using the Wick
theorem:

\begin{eqnarray}\label{T2nT2n}
\langle\tilde{T}_{2n}(z)\tilde{T}_{2n}(w)\bar{\xi}\xi\rangle=-
\frac{2^{2n-2}\{(2n-1)!\}^2}{(z-w)^{4n}}.
\end{eqnarray}
that as compared with
$\langle\tilde{T}_{2n}(z)\tilde{T}_{2n}(0)\bar{\xi}\xi\rangle=\frac{c_{2n}}{2z^4}$,
concludes the previous result (\ref{c-2n}).

\vspace{1cm}
{\Large {\bf Acknowledgment}}\\

We would like to thank M.Gaberdiel for useful comments, and
S.Moghimi-Araghi for reading the manuscript and helpful comments.

\vspace{1cm}
\begin{appendix}

\section{Some OPEs  and Correlations
}\label{appdisk}\setcounter{equation}{0}\

In this appendix, we calculate some of OPEs mentioned in sections
(\ref{Logarithm}) and (\ref{Sugawara}). These explicit forms may
be used in calculating RG equations. Moreover, using these
expressions one can simply read off the two point functions of
the fields or their derivatives.\\ The OPE of field $\psi$ with
itself is rigorously obtained using Wick theorem as follows :
\begin{eqnarray}\label{sisi}
\psi(z)\psi(w)&=&-\frac{:\theta\bar{\theta}:(w)}{2|z-w|^4}(1+\log|z-w|^2)+\frac{I}{8|z-w|^4}\{2(1+\log|z-w|^2)
+\log^2|z-w|^2\}\nonumber\\&-&\frac{\bar{\partial}:\theta\bar{\theta}:(w)}{4|z-w|^2(z-w)}(1+\log|z-w|^2)\nonumber\\&+&
\{\frac{\bar{T}(w)}{8(z-w)^2}
+\frac{T(w)}{8(\bar{z}-\bar{w})^2}\}(4+\log|z-w|^2-\log^2|z-w|^2)\nonumber\\&-&\frac{\bar{\partial}(\theta\bar{\partial}\bar{\theta}
+\bar{\partial}\theta\bar{\theta})}{8(z-w)^2}(2+\log|z-w|^2)-\frac{\partial(\theta\partial\bar{\theta}
+\partial\theta\bar{\theta})}{8(\bar{z}-\bar{w})^2}(2+\log|z-w|^2)-
\frac{\psi(w)}{|z-w|^2}\nonumber\\&+&\frac{\phi(w)}{4|z-w|^2}
(2+\log|z-w|^2)-\{\frac{\bar{t}(w)}{2(z-w)^2}+\frac{t(w)}{2(\bar{z}-\bar{w})^2}\}(1-\log|z-w|^2)
\nonumber\\&-&\frac{4\bar{\partial}\psi(w)-\partial
\bar{t}(w)\log|z-w|^2}{4(z-w)}+
\frac{\bar{\partial}\phi(w)}{16(z-w)}(2+\log|z-w|^2)\nonumber\\&-&\frac{4\partial\psi(w)
-\bar{\partial}t(w)\log|z-w|^2}{4(\bar{z}-\bar{w})}+\frac{\partial\phi(w)}{16(\bar{z}-\bar{w})}(2+\log|z-w|^2)
+\cdot\cdot\cdot
\end{eqnarray}
Moreover, the OPE of $\psi$ with the logarithmic partner of
identity i.e. $:\theta\bar{\theta}:$, turns out to be :
\begin{eqnarray}\label{sitete}
:\theta\bar{\theta}:(z)\psi(w)&=&-\frac{:\theta\bar{\theta}:(w)}{2|z-w|^2}
-\frac{\bar{\partial}:\theta\bar{\theta}:(w)}{4(z-w)}\log|z-w|^2
-\frac{\partial:\theta\bar{\theta}:(w)}{4(\bar{z}-\bar{w})}\log|z-w|^2
\nonumber\\&+&\log|z-w|^2\psi(w)-\frac{1}{4}\log^2|z-w|^2\phi(w)+
\cdot\cdot\cdot.
\end{eqnarray}
Since all fields have zero expectations except
$:\theta\bar{\theta}:$, correlation of $\psi$ with itself and
$:\theta\bar{\theta}:$ can be simply obtained :
\begin{eqnarray}\label{Apensitete}
\langle\psi(z)\psi(w)\rangle&=&\frac{1}{2|z-w|^4}\{1+\log|z-w|^2\},\nonumber\\
\langle\psi(z):\theta\bar{\theta}:(w)\rangle&=&\frac{1}{2|z-w|^2}.
\end{eqnarray}
Which are consistent with the equations (\ref{csisi2}) and
(\ref{csisi3}).\\

We have also calculated the OPEs of the logarithmic partner of
energy-momentum tensor $t$ with the fields
$:\theta\bar{\theta}:$, $\phi$ and $\psi$.\\ The first OPE is :
\begin{eqnarray}\label{ttete3}
:\theta\bar{\theta}(z):t(w)&=&-\frac{:\theta\bar{\theta}:(w)}{2(z-w)^2}
-\frac{1}{2(z-w)}\log|z-w|^2\partial:\theta\bar{\theta}:(w)\nonumber\\&-&\frac{1}{4}\log^2|z-w|^2T(w)+\log|z-w|^2
t(w)+\cdot\cdot\cdot.
\end{eqnarray}
According to this OPE, the equation (\ref{ttetta}) can be clearly
obtained.\\
The OPE of $t$ with $\phi$ has the following form :
\begin{eqnarray}\label{ttete1}
t(z)\phi(w)&=&-\frac{\partial:\theta\bar{\theta}:(w)}{2(z-w)^2(\bar{z}-\bar{w})}
-\frac{\partial(\partial\theta\bar{\theta}+\theta\partial\bar{\theta})}{2|z-w|^2}\nonumber\\&+&\frac{2\psi(w)
-\phi(w)}{2(z-w)^2}-\frac{\partial\phi(w)-2\partial\psi(w)+\bar{\partial}t(w)}{2(z-w)}-\frac{7}{8}\frac{\partial
T(w)}{(\bar{z}-\bar{w})} +\cdot\cdot\cdot.
\end{eqnarray}
Note that the corresponding correlation vanishes.\\

Finally, the most singular terms appearing in the OPE of $t$ with
$\psi$ are :

\begin{eqnarray}\label{ttete2}
t(z)\psi(w)&=&\frac{1}{4(z-w)^3(\bar{z}-\bar{w})}\{I+\log|z-w|^2I-2:\theta\bar{\theta}:(w)\}\nonumber\\
&+&\frac{1}{4(z-w)^3}\{1+\log|z-w|^2\}\bar{\partial}:\theta\bar{\theta}:(w)\nonumber\\
&-&\frac{1}{4(z-w)^2(\bar{z}-\bar{w})}
\{1+\log|z-w|^2\}\partial:\theta\bar{\theta}:(w)+\cdot\cdot\cdot,
\end{eqnarray}
Which leads to the two point function
\begin{eqnarray}\label{Appenttete2}
\langle t(z)\psi(w)\rangle=\frac{1}{2(z-w)^3(\bar{z}-\bar{w})}.
\end{eqnarray}

\end{appendix}

\end{document}